\documentclass[twocolumn]{autart}    

\usepackage{amssymb}
\usepackage{indentfirst}
\usepackage{hyperref}
\usepackage{graphicx}          
\usepackage{epstopdf} 
\usepackage[brazilian]{babel} 
\usepackage[utf8]{inputenc}
\usepackage{multicol}
\usepackage{amsmath} 
\usepackage{enumerate} 
\usepackage{booktabs} 
\usepackage{array}
\usepackage{rotating} 
\usepackage{multirow} 
\usepackage{float}

\begin{document}

\begin{frontmatter}

\title{Evidence of chaos and nonlinear dynamics in the Peruvian financial market}


\author[Paestum]{Alexis Rodriguez Carranza}\ead{arz@upn.edu.pe},    
\author[Baiae]{Marco A. P. Cabral}\ead{mapcabral@ufrj.br}  
\author[Rome]{Juan C. Ponte Bejarano}\ead{juan.ponte@upn.edu.pe}.               
\address[Paestum]{Department of Sciences - Private University of the North, Perú}  
\address[Baiae]{Federal University of Rio de Janeiro, Brazil}        
\address[Rome]{Department of Sciences - Private University of the North, Perú}             

\begin{keyword}                           
Caos; Dinâmica não lineal; Séries temporais.               
\end{keyword}                             

\begin{abstract}              
Physicists experimentalists use a large number of observations of a phenomenon, where are the unknown equations that describe it, in order to play the dynamics and obtain information on their future behavior. In this article we study the possibility of reproducing the dynamics of the phenomenon using only a measurement scale. The Whitney immersion theorem ideas are presented and generalization of Sauer for fractal sets to rebuild the asymptotic behaviour of the phenomena and to investigate, chaotic behavior evidence in the reproduced dynamics. The applications are made in the financial market which are only known stock prices 

\end{abstract}

\end{frontmatter}

\section{Introduction}
For a study of the asymptotic behavior of solutions of a system, the area of dynamical systems has developed a lot of tools , but in many phenomena as financial markets, equations that model them are unknown and the only available information is a temporal set of measures.

\begin{figure}
\begin{center}
\includegraphics[height=6cm]{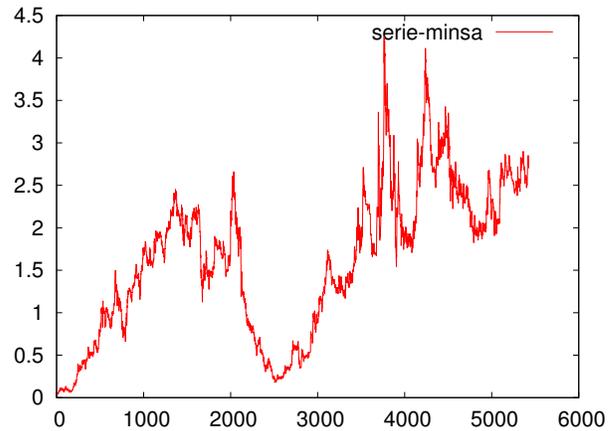}    
\caption{Evolution of stock prices of the MINSUR company}  
\label{fig1}                                 
\end{center}                                 
\end{figure}

A time series is a function $s:I\!\!R \rightarrow I\!\!R$, and its image could be considered as observations taken overtime in a particular phenomenon. Those pieces of observations have information about the system and we are ready to answer questions like does that serie have enough information to rebuild the dynamics? If it is so, can we predestine its future behaviour?. But there is one more general question. Will the financial market have a random behaviour or not?. The chaos theory could give us an answer where systems with  an apparently ramdom behaviour, comes from a deterministic system, it means a completely modeling system by equations. Chaotic system complexity lies primarily in sensitivity to initial conditions, small variations produce big changes.This sensitivity is quatified by Lyapunov exponents. We will do our research in a time series of the stock prices in the mining company MINSUR.SA to determine concrete evidence of chaos presence in that series.
Graphic(\ref{fig1}) shows evolution of stock prices of the company.

\section{Teoría do Caos}
\begin{figure}[H]
\begin{center}
\includegraphics[height=6cm]{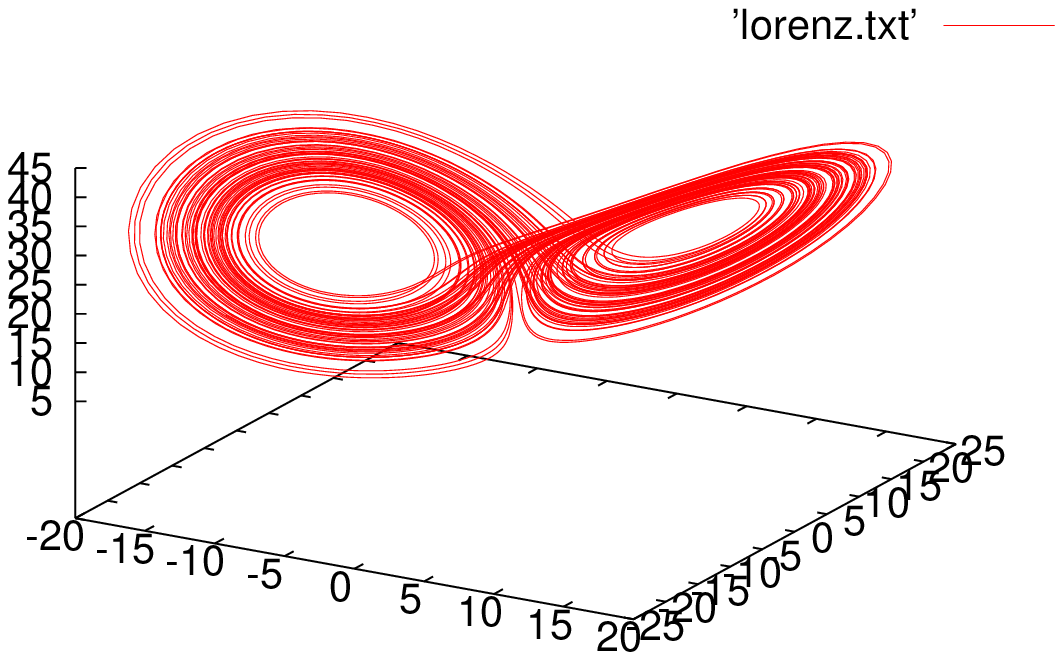}    
\caption{Sistema de Lorenz}  
\label{fig2}                 
\end{center}                                 
\end{figure}

To show the use of chaos theory in time series, we consider the Lorenz system, which is given by:
\begin{eqnarray*}
\frac{dx}{dt} &=& 10(y-x) \\
\frac{dy}{dt} &=& x(b-z) - y \\
\frac{dz}{dt} &=& xy -cz 
\end{eqnarray*}
The dynamics of system solutions is shown in the figure(\ref{fig2}) 

\begin{figure}[H]
\begin{center}
\includegraphics[height=6cm]{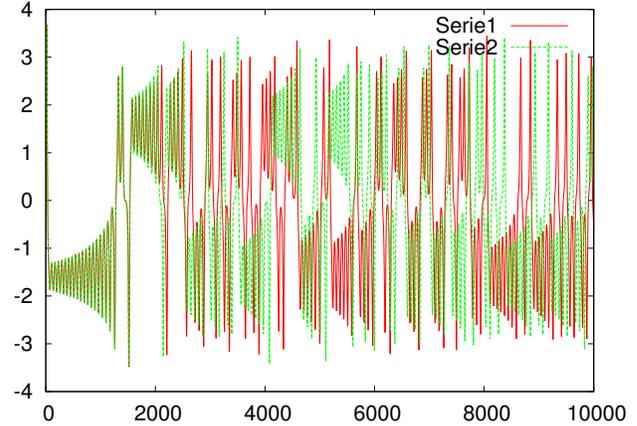}    
\caption{Component x of Lorenz system for close initial conditions}  
\label{fig3}                                 
\end{center}                                 
\end{figure}

It is well known Lorenz system is sensitive to small variations with initial data, it means, solutions with very close initial data is not exact exponentially\cite{mane}. Figure(\ref{fig3}) shows such a behaviour in the first component of solutions for very close data initial. The series in the figure(\ref{fig3}) were generated with initial data in points $(1,2,3)$ and $(1.00001,2.00001,3.00001)$ in time with variation in mili seconds. Notice the trajectories differ after 3 seconds, it means, small variations generate big changes quickly. So that, chaos theory could explain some phenomena in the financial market, where variations like speculations generate changes in a short term. One of the goals in this work is to investigate chaotic behavior on time series in the financial market precisely applied in the Peruvian financial system

The main task to study the chaotic behavior in a time series is the rebuilding of the dynamics it has. If a time series comes from a deterministic system, so that it has dependence between its components given by the system of equations that modeling and contains geometric information considering that trajectories converge to an attractor which will be inmerse in some Euclidian space. Using such information and The Whitney immersion theorem ideas in which a dimensional compact manifold $n$ can be inmersed in $I\!\!R^{2n}$ and its generalizations for fractal sets given for Tim Sauer\cite{sauer}, we will rebuild the dynamics knowing the information of one of the components of Lorenz system only. 

Now we can see the probabilistic version of Theorem of Withney\cite{sauer}
\section{Main theorems}
\begin{thm}{\label{preva}}(Prevalent theorem of Whitney)
If $A$ is a smooth compact manifold  of dimension  $d$ contained on $I\!\!R^{k}$. Almost all smooth map, $F: \!\!R^{k}\rightarrow \!\!R^{2d+1}$, is an immersion of $A$.
\end{thm}

The theorem(\ref{preva}) says it is possible to rebuild the existing dynamics with projections in Euclidian spaces. 
Despite we have the theorem of Whitney, we  still have a practical problem. We only can get a temporal observation of the dynamics in the financial market, the evolution of the stock prices and we will need $2d+1$ 1observations of the phenomenon. So, the obtained results are not enough. Takens assumed this problem adding the contained dynamics in the time series in the Theorem of Whitney. It is a projection via a function of observation of the phase space where the dynamical system is developed for $I\!\!R$. Therefore it contains information about the dynamics. For this, it defines the {\it delay cordinates}, which only need a temporary observation.

\begin{defn}{\label{delacor}}
If $\Phi$ is a dynamics system over a manifold $A$, $T$ a positive integer (called of {\bf \it delay}), and $h:A\rightarrow  \!\!R$ a smoth function. We defines the map of delaying coordinates $F_{(\Phi,T,h)}: A\rightarrow \!\!R^{n+1}$ 
\[F_{(\Phi,T,h)}(x)=(h(x),h(\Phi_{T}(x)),h(\Phi_{2T}(x)),\dots,h(\Phi_{nT}(x))) \]
\end{defn}

Taken\cite{Takens} demonstrated a new version of theorem of Whitney for the delaying coordinates

\begin{thm}(Takens)
$A$ is a dimensional compact manifold $m$, $\{\Phi_k\}_{k\in \mathbb{Z}}$ a discrete dynamical system over $A$ generated by $F: A\rightarrow A $, and a function of classe $C^2$ $h:A \rightarrow I\!\!R$.
is a generic characteristic that of map $F_{(\Phi,h)}(x): A \rightarrow I\!\!R^{2m+1} $ defined by 
\[F_{(\Phi,h)}(x)=(h(x),h(\Phi_k(x)),h(\Phi_{2k}(x)),\dots,h(\Phi_{nk}(x))) \]
is an immersion.
\end{thm}

The final generalization used in this article was given by B.Hunt, T. Sauer and J. Yorke\cite{Embedo}, that is a fractal version of theorem of Whitney for the delaying maps set with $A$ being a fractal set.

\begin{thm}{\label{delafraw}}(Fractal Delay Embedding Prevalence Theorem)
$\Phi$ A dynamics system over an open subset  $U$ of $I\!\!R^{k}$, and $A$ a compact subset of $U$ with box dimension $d$. $n>2d$ an integer and $T>0$. Asume that $A$ contains only a
finite number of equilibrio points, it does not contains periodic orbits of $\Phi$ of period $T$ o $2T$,
a finit number of periodic orbits of $\Phi$ of period $3T$, $4T,\dots,nT$, and these periodic orbits linearizations have different eigenvalues. So for almost every smooth function(in the sense prevalente) smooth function $h$ over $U$, the delay coordinates map $F_{(\Phi,T,h)}: U\rightarrow \!\!R^{n}$ is injective over $A$
\end{thm}
The theorem(\ref{delafraw}) does not provide an estimative about the smaller dimension for which almost every delaying coordinates map is injective. However there are numerical algorithms which allow calculate the immersion dimension and the delaying time in the reconstruction.Following, we show the reconstruction of the dynamics generated for the system of Lorenz using only the coordinate $x$ of the system

\section{4.	Examples of reconstruction of the attractor using delay coordinates}\label{subsec:expolyap}
we use the Lorenz's attractor to show the technique of delaying coordinates. The function of observation $h$, was the projection in the $x$ axis.
\[
\begin{array}{cccc}
  h :& I\!\!R^3 & \rightarrow & I\!\!R \mbox{  }\\
   & (x,y,z) & \rightarrow & h(x,y,z)=x
\end{array}
\]
The time series will be formed by $x$ coordinates of the trajectories that are numerical solutions of the equation. Fixed the dimension, $n=3$, we change the value of T $T$. According to Lorenz's  we use delaying times, $T=1$, $T=10$, $T=17$, $T=90$. In the case $T=1$ points in the space are highly correlated and the graphic is almost a straight line. At the other extreme, $T=90$, points are not correlated and the gotten graphic does not represent the reconstruction of the attractor. The optimal  delay time was $T=17$. 

\begin{figure}[H]
\centering
\begin{minipage}[c]{4cm}
\includegraphics[width=5cm]{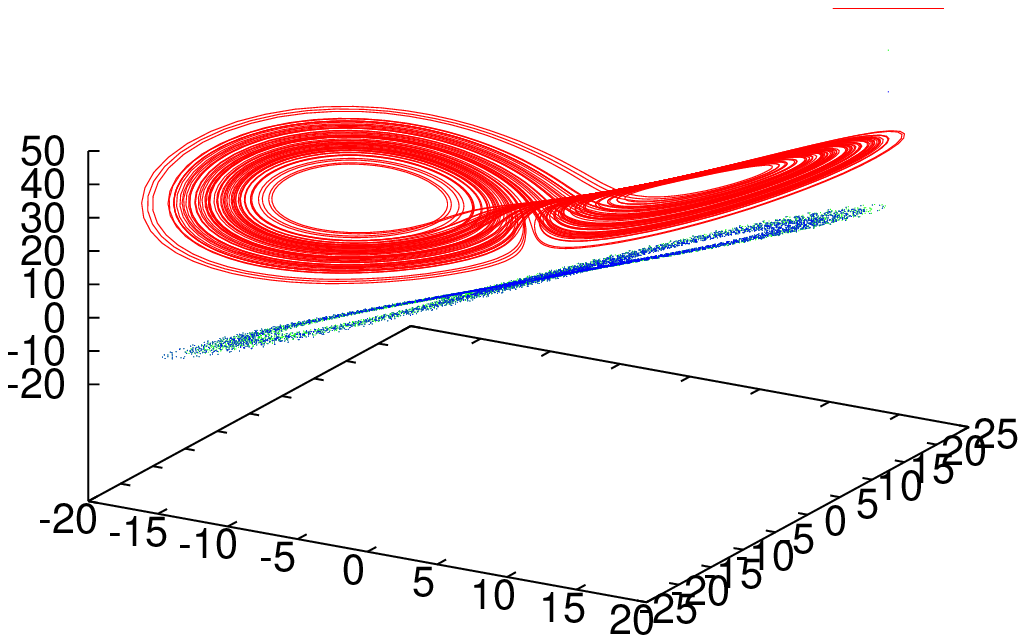}
\end{minipage}
\begin{minipage}[r]{4cm}
\includegraphics[width=5cm]{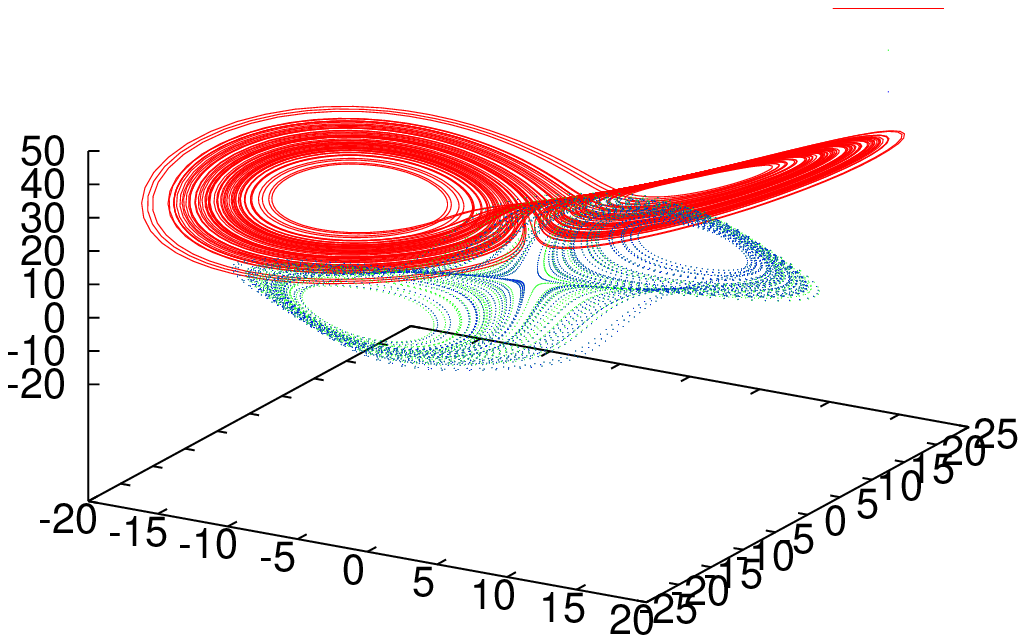}
\end{minipage}
\\
\begin{minipage}[c]{4cm}
\includegraphics[width=5cm]{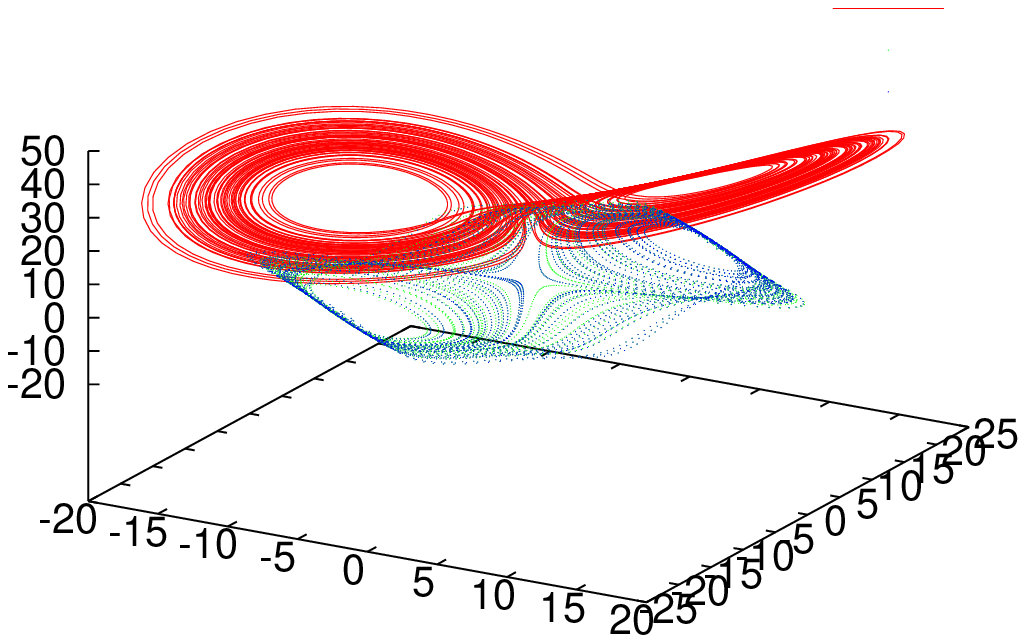}
\end{minipage}
\begin{minipage}[r]{4cm}
\includegraphics[width=5cm]{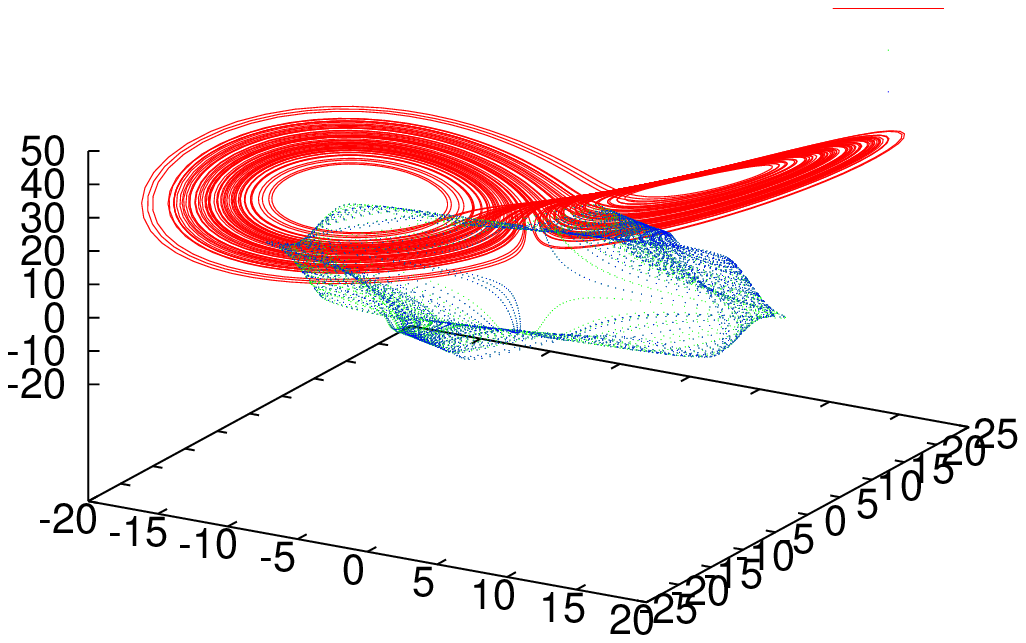}
\end{minipage}

\caption{reconstruction of attractor using proyections with time delay d=1, 10, 17 and 90 respectively}\label{figures}
\end{figure}

\begin{figure}[H]
\begin{center}
\includegraphics[height=6cm]{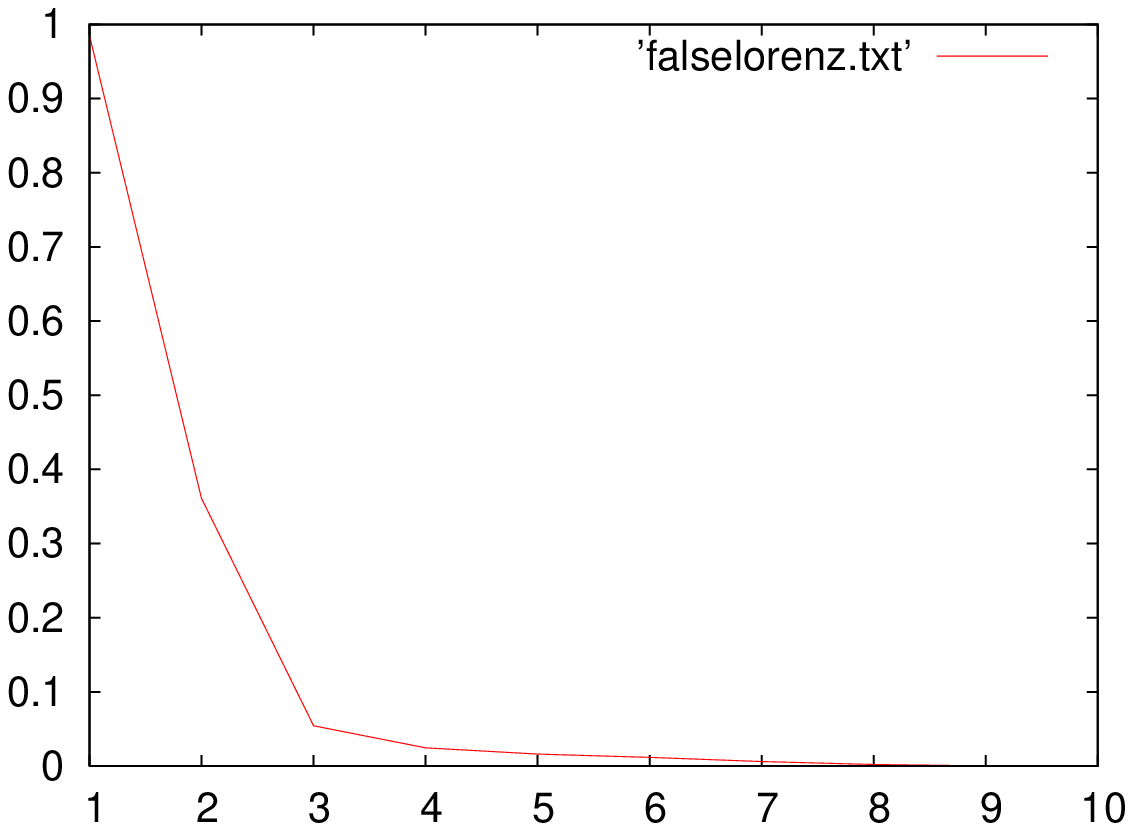}    
\caption{Dimension of inmersión}  
\label{fig10}                                 
\end{center}                                 
\end{figure}

To determine the dimension where is rebuilt the dynamics, the technique of the neighbours is used. The idea is about setting a dimension and calculate the distance between close points, if the attractor does not have enough freedom, so when we raise the dimension of the space, those next points get separeted and are pointed as false neighbours. The procedure goes on until the percentage of false neighbours is small. In the series formed with the first components of Lorenz's system, the number of false neighbours is almost zero when the dimension of immersion is $3$, figure(\ref{fig10}).

\section{Identifying non linearity and chaos}
First of all, we will give the definition of chaos according to Devaney\cite{Devaney}.
\begin{defn}{\label{defcaos}}
If $V$ is an interval. We say that $f:V\rightarrow V$ is chaotic over $V$ if:
\begin{itemize}
    \item[1.] $f$ has sensibility to initial conditions;
    \item[2.] $f$ is transitive;
    \item[3.] The periodic points are dense in $V$.
\end{itemize}
\end{defn}
We say that $f$ has sensibility to initial conditions when the solutions with initial conditions very close, diverge quickly.
\begin{defn}
A map $f:[0,1]\rightarrow [0,1]$ has sensitive dependence in $x$ if the next condition holds for some $\delta>0$: for each open I on $[0,1]$, containing $x$, there is a $y\in I$ and $n\in I\!\!N$ such that
\[|f^{n}(x)-f^{n}(y)|\geq\delta.\]
\end{defn}

\begin{defn}
A map $f:[0,1]\rightarrow [0,1]$ is transitive if for each pair of subintervals $I$ y $J$ de $[0,1]$ there is $n$ such that  $f^{n}(I)\cap J\leq \phi$
\end{defn}
Intuitively, $f$ is transitive implies the orbit
of an interval $I_0$ is dense on $[0,1]$.

Given chaos definition, we will investigate the existence of characteristics $1)-3)$ in a time series. In this article we are interested in searching the characteristic $1$ which
is characterized by Lyapunov exponents.

It is important to know lineal systems can not be chaotic. We will identify no lineal dependences on data. The most used techniques are: method of surrogates data and the test BDS.\\
Kantz e Schreiber\cite{Livroseries} recommend that before doing any non lineal analysis on a data set, it is good practice to check if there is no linearity. Because the linear systems are not chaotic and the linear stochastic processes can produce very complex series which can be misunderstood as a product of chaotic system. We presented the technique of surrogate data and  the BDS test.

\subsection{Surrogate data}
A very useful technique used by physicist is surrogate generation of {\it Surrogate data}, a procedure done by Theiler, Eubank, Longtin, Galdrikan y Farmer\cite{Surro}.
From the original data , it generates a set of random series so that, these keep the linear properties of the original series (average, variance , Fourier spectrum) but eliminating the possible nonlinear dependencies.Then an indicator is evaluated, which is sensitive to nonlinear dependencies and try to reject the null hypothesis, which states that data are obtained by a stochastic linear process. If the null hypothesis is true, then the substitute series procedure would not affect the indicator

The most used indicators are correlation dimension. It was used by Small et. al.\cite{small} and no linear prediction which was used by Kaplan\cite{Kaplan}.

We will illustrate the technique of substitute data applied in Lorenz series and use the nonlinear prediction.

The election was due to deterministic systems the prediction on a short term is possible, in contrast to Stochastic systems in which the prediction in a short term is impossible . For the prediction nonlinear we will use the method developed by Hegger, Kantz and Schreiber\cite{Heger}, where the value of predicting is the average of ``future'' values. This state $n+k$, is defined by average of the closed values. 
It is calculated by the following equation
\[\overline{S}_{n+k}=\frac{1}{|U_n|}\sum_{S_{j}\in U_n}S_{j+k},\]

Where $U_n$ is the neighborhood, $S_j$ are values within $U_n$ and $S_{j+k}$ are the values at time $n+k$. held the
prediction, this is compared with the quadratic average error of prediction on future value of the original serie with the surrogate series. If the prediction error of the series original is smaller than all the surrogate series, then we reject the null hypothesis. This graphic shows the results.
\begin{figure}[H]
\centering
\includegraphics[height=.25\textheight]{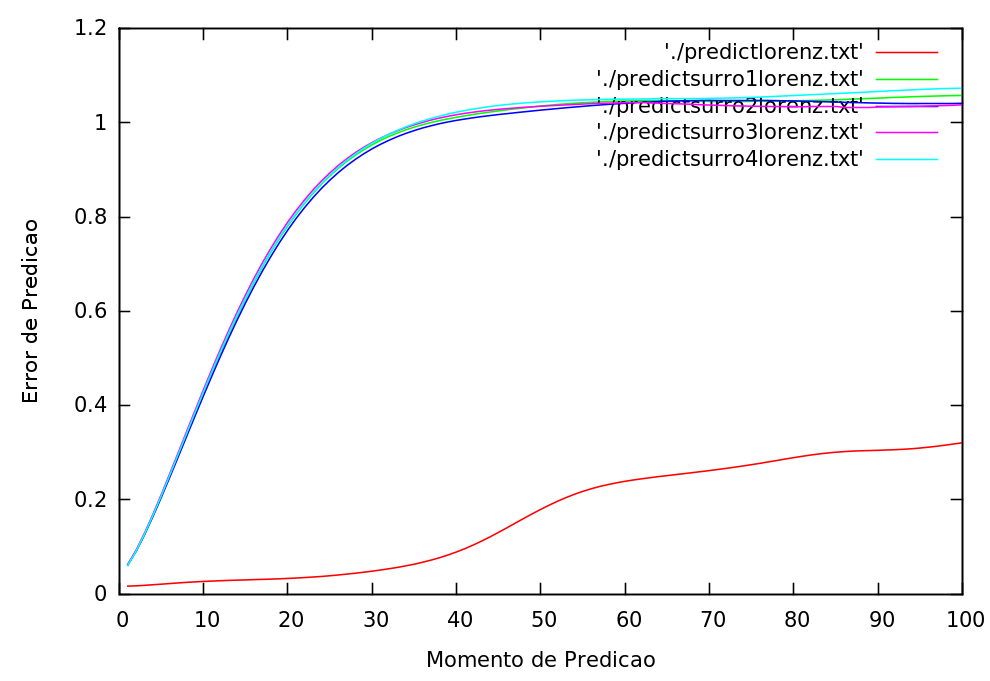}
\caption{ In the graphic the proyection errors of Lorenz series and its surrogate series}
\end{figure}

The projection errors of the original series were smaller than its substitutes. Like this, we reject the null hypothesis.

The following technique is more about statistics.

\section{BDS Test}
It was developed by Brock, Dechert and Scheinkman\cite{BDS}, It's based on the correlation dimension to detect nonlinear structure in a time series. Additionally this test can be used to test how good is the fit estimation model

Given a time series $s_1,s_2,\dots,s_N$, Using the method of delay coordinates, we have
$M=N-(m-1)\tau$ vectors in $I\!\!R^m$, $X_{i}=(s_i,s_{i+\tau},\dots,s_{i+(m-1)\tau})$, where
$\tau$ is the delay time, $m$ is the immersion dimensión.
The correlation dimension is given by the formula,
\begin{eqnarray*}
C(M,r) &=& \frac{1}{M(M-1)}\sum_{i=1}^{M}M_{X_{i}^{m}}(\epsilon) \\
&=&\frac{1}{M(M-1)}\sum_{i\neq j}\theta(\epsilon-||X_{i}-X_{j}||),
\end{eqnarray*}
Where $M_{X_{i}}(\epsilon)$ is the function point mass, which indicates the probability where the points $X_i,X_j$ are $\epsilon$ close to each other. 

For practical and didactic issues we consider $\tau=1$. Now if $m=2$ we will get 
\begin{eqnarray*}
X_{i}&=&(x_{i},x_{i+1}),\\
X_{j}&=&(x_{j},x_{j+1}).
\end{eqnarray*}
Then, $||X_{i}-X_{j}||\leq\epsilon$ implies $|x_{i}-x_{j}|\leq\epsilon$ and
$|x_{i+1}-x_{j+1}|\leq\epsilon$.

This allows to say that if the points $X_{i}$ e $X_{j}$ are closing to each other, then, the points from series $x_{i}$ and
$x_{j}$ as well. It happens the same with points $x_{i+1}$ and $x_{j+1}$. Thus,
\begin{eqnarray*}
M_{X_{i}}(\epsilon)&=&\mbox{P}(||X_{i}-X_{j}||\leq\epsilon)\\&=&
\mbox{P}(|x_{i}-x_{j}|\leq\epsilon;|x_{i+1}-x_{j+1}|\leq\epsilon)
\end{eqnarray*}
Where $P$ indicates the probability. If the data $x_1,x_2,\dots,x_N$ are IID(independent and identically distributed), then $P(|x_{i}-x_{j}|\leq\epsilon)=P(|x_{i+1}-x_{j+1}|\leq\epsilon)$ and we have,
\begin{eqnarray*}
M_{X_{i}}(\epsilon)&=&\mbox{P}(|x_{i}-x_{j}|\leq\epsilon)
\mbox{P}(|x_{i+1}-x_{j+1}|\leq\epsilon)\\
&=& \mbox{P}^2(|x_{i}-x_{j}|\leq\epsilon).
\end{eqnarray*}
Thus, in dimension $m$ we have \[M_{X_{i}}(\epsilon)=\mbox{P}^m(|x_{i}-x_{j}|\leq\epsilon)\]
Then,
\begin{eqnarray*}
C_m(M,\epsilon)&=&\frac{1}{M}\sum_{i=1}^{M}M_{X_{i}}(\epsilon)\\
&=&\frac{1}{M}\sum_{i=1}^{M}P^m(|x_{1}-x_{2}|\leq\epsilon)\\
&=& P^m(|x_{1}-x_{2}|\leq\epsilon).
\end{eqnarray*}
As $P(|x_{1}-x_{2}|\leq\epsilon)=P(|x_{1}-x_{3}|\leq\epsilon)=\dots=P(|x_{i}-x_{j}|\leq\epsilon)=\dots$.
This suggests that the data is IID, then,
\[C_{m}(M,\epsilon)=C_{1}^{m}(M,\epsilon),\]

Where $C_{m}(M,\epsilon)$ is the integral of correlation in dimension $m$ and $C_{1}(M,\epsilon)$ is the integral of correlation in dimension one.

In, Brock, Dechert y Scheinkman\cite{BDS}, statistical BDS test is defined by:
\[\sqrt{M}\frac{C_{m}(M,\epsilon)- C_{1}^{m}(M,\epsilon)}{\sigma_{m}(\epsilon)}\]

Where $\sigma_{m}(M,\epsilon)$ is the estimation of the standard asymptotic error: $C_{m}(M,\epsilon)- C_{1}^{m}
(M,\epsilon)$. They proved that:
\[\sqrt{M}\frac{C_{m}(\epsilon)- C_{1}^{m}(\epsilon)}{\sigma_{m}(\epsilon)}\sim N(0,1)\]

Introducing to BDS test applied in Lorenz series and a series consisting of random numbers.
\subsection{BDS Test for Lorenz series}
\vspace{1cm}
\begin{center}
\begin{tabular}{cc} \hline
Dimension of immersion(m) & BDS \\
\hline
1 &   165.270466  \\
2 &   447.623951  \\
3 &   578.535241  \\
4 &   772.617763  \\
5 &   1078.467687 \\
6 &   2338.679429  \\
7 &   3593.653394  \\
8 &   5640.618871  \\ \hline
\multicolumn{2}{l}{}
\end{tabular}
\end{center}

\subsection{Test BDS for a series consisting of random numbers}
\vspace{1cm}

\begin{center}
\begin{tabular}{cc} \hline
Dimension of immersion(m) & BDS \\
\hline
1 & 0.560594  \\
2 & 0.560594   \\
3 & 0.425293   \\
4 & 0.069765  \\
5 & 0.116118 \\
6 & 0.336857  \\
7 & 0.633835  \\
8 & 0.524584  \\ \hline
\multicolumn{2}{l}{}
\end{tabular}
\end{center}
\vfill

To understand the results, let's start talking about the degree of significance of a hypothesis. To ask a null hypothesis, in the case of the BDS test the null hypothesis is that the observations are independent and identically distributed (IID), it can be make a mistake of rejecting the null hypothesis being true.
Suppose that the probability of committing this error is $\alpha$, namely:
\[\alpha=P(\mbox{rejeitar}\quad H_0|H_0\,\, \mbox{é}\quad\mbox{verdadeiro})\]
$\alpha$ is called the degree of significance of this mistake. Now, in the BDS test:
\begin{eqnarray*}
H_0: C_{m}(M,\epsilon) &=& C_{1}^{m}(M,\epsilon)\\
H_a: C_{m}(M,\epsilon) &\neq& C_{1}^{m}(M,\epsilon)
\end{eqnarray*}
By the test BDS, know that
\[\sqrt{M}\frac{C_{m}(M,\epsilon)-C_{1}^{m}(M,\epsilon)}{\sigma_{m}(\epsilon)}\sim N(0,1)\]
Then,
\[
\quad\alpha =P(\mbox{rejeitar}\quad H_0|H_0\,\, \mbox{é}\mbox{verdadeiro})
\]
\[
\alpha = P(|\sqrt{M}\frac{C_{m}(M,\epsilon)-C_{1}^{m}(M,\epsilon)}{\sigma_{m}(\epsilon)}|>z_c |H_0\,\, \mbox{é}\quad\mbox{verdadeiro})
\]
\[
\alpha = P(|\sqrt{M}\frac{C_{m}(M,\epsilon)-C_{1}^{m}(M,\epsilon)}{\sigma_{m}(\epsilon)}|>z_c \sim N(0,1))
\]
\[
\alpha = 1-2P(0\leq \sqrt{M}\frac{C_{m}(M,\epsilon)-C_{1}^{m}(M,\epsilon)}{\sigma_{m}(\epsilon)}\leq z_c)
\]
So,
\[
P(0\leq \sqrt{M}\frac{C_{m}(M,\epsilon)-C_{1}^{m}(M,\epsilon)} {\sigma_{m}(\epsilon)}\leq z_c) = \frac{1-\alpha}{2}
\]

If we look at the tables of the normal distribution $N(0,1)$, we see that you to a degree of signicancia of $\alpha=5\%$, $z_c =1.96$. As, we reject the null hypothesis, with degree of signicancia of $5\%$, if $|\sqrt{M}\frac{C_{m}(\epsilon)-C_{1m}(\epsilon)}{\sigma_{m}(\epsilon)}|>1.96$. Noting again the tables we can observe that in the case of Lorenz series, we reject the null hypothesis, the data are IID. Otherwise case series formed by random numbers.

Detected non-linealidade in the time series, the next step is to search sensitive dependence on initial conditions.

\subsection{Lyapunov exponents}\label{subsec:expolyap}

This section we obtain results that characterize the sensitive dependence on initial conditions. The study of the attractor and the dynamics is performed in the space reconstructed using delay coordinates.

This section we obtain results that characterize the sensitive dependence on initial conditions. The study of the attractor and the dynamics is performed in the space reconstructed using delay coordinates.
Chaotic systems are, generally, strange attractors which are generated by stretching and contractions.
Stretches are responsible for sensitivity on initial conditions and are characterized by the exponential distance in finite time, paths with initial conditions very next. Contractions allow the pathlines may not move away indefinitely and that they remain confined to a bounded phase space region. Distance speed is measured by the Lyapunov exponents.

In the case of a discrete dynamic defined by a map system $f:\!\!R^{n}\rightarrow \!\!R^{n}$,
\[
T(x)=D_x f
\]
is the Jacobian matrix of $f$. then writing
\[
T(x)=T^{n}_{x}=T(f^{n-1}x),\dots,T(fx),T(x)
\]
The largest Lyapunov exponent is given by:
\[
\lambda_{1}=\lim_{n\rightarrow\infty}\frac{1}{n}|\!|T^{n}_{x} u|\!|
\]

for almost all vector $u$, for more details see\cite{Ergodic}. As we can see the first problem for the calculation of Lyapunov exponents is to obtain the matrizes $T(x)=D_x f$. In the case of having only a time series the calculation of Lyapunov exponent is even more complicated. Eckman y Ruelle\cite{Ergodic},\cite{EckLyap} proposed the first algorithms based on the following ideas.

Consider a dynamic generated by the differential equation:
\begin{equation}{\label{sisnolinear}}
\frac{dx(t)}{dt}=F(x(t))
\end{equation}
In  $I\!\!R^{n}$. If we do $u(t)=F(x(t))$ in the previous equality we get,
\begin{equation}{\label{sislinear}}
\frac{du(t)}{dt}=(D_{x(t)}F)(u(t)).
\end{equation}

The retrieved system is linear in $u$. Then, if the system(\ref{sisnolinear}) solutions are $x(t)=f^{t}(x(0))$, then the solutions of the system(\ref{sislinear}) are $u(t)=(D_{x(0)}f^{t})u(0)$.
So the matrix, $T_{t}^{x}=D_{x}f^{t}$, can be calculated   integrating the equations (\ref{sisnolinear}).
This idea for a linear system to calculate the matrix, $T_{t}^{x}=D_{x}f^{t}$, was used by Eckman and Ruelle\cite{EckLyap} for estimating of Lyapunov exponents from a time series.

Consider a time series, $s_1,s_2,\dots,s_n$. Using delay coordinates, obtain points
$X_1, X_2,\dots, X_m$, in the space $m$-dimensional of reconstruction. If we develop $f$ around point
$x_i$, we would have that for $x_j$ next to it,
\begin{eqnarray}{\label{separaexpo}}
f(X_j) &\approx & f(X_i)+Df_{X_i}(X_j -X_i)\nonumber \\
f(X_j)-f(X_i) &\approx & Df_{X_i}(X_j-X_i)
\end{eqnarray}

So, $T_{\tau}^{X(i)}=D_{X(i)}f^{\tau}$ is estimated by the best linear approximation of the map that for points $x(j)$ next to point $X(i)$, takes $X(j)-X(i)$ in points $f^{\tau}(X(j))-f^{\tau}(X(i))= X(j+\tau)-X(i+\tau)$ close. This process can be thought as the linear approximation of the derivatives at each point in space lathe overhauled do. With the same ideas, authors such as Wolf, Swift, Swinney Vastano\cite{Wolf}, M. Sano, y. Sawada\cite{SanoWa} proposed an algorithm for the calculation of Lyapunov exponents.

Another factor to consider is that the Lyapunov exponents are defined in terms of limits, which guarantees its invariance by smooth transformations. If we think of a finite series, there is no guarantee that interference (noise) data, lead to the same exponents. Rosenstein, Collins and De Luca\cite{Lyapr} and Holger Kantz\cite{Lyapk} directly presented new techniques based on the idea that the greatest exponent of Lyapunov measure the average speed of estrangement of the paths with conditions next. The estimation of the maximum exponent of Lyapunov using the algorithm of Rosenstein was made in the reconstructed attractor of Lorenz. The value of $0.7973003$ was obtained. 
\begin{figure}[H]
\centering
\includegraphics[height=.25\textheight]{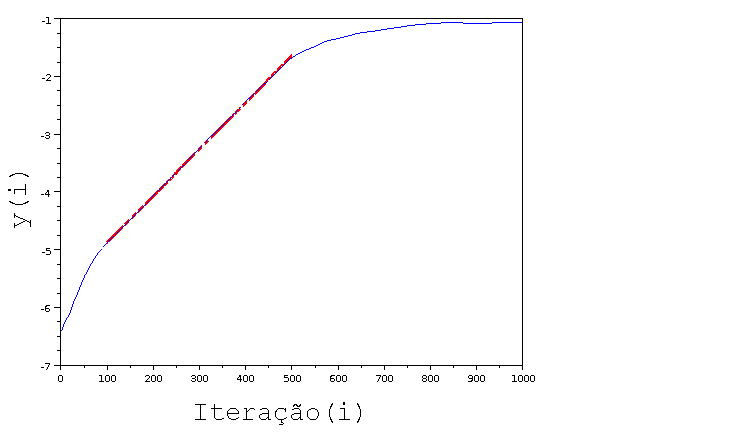}
\caption{Departure of paths with initial conditions close to Lorenz attractor.}
\end{figure}
\section{Applications}
The series studied was the series of prices of the shares of the mining company Minsur, counting on prices from $13/10/1993$ up to  $26/11/2014$ with a total of $21$ years of daily observations. 
\subsection{Identifying non-linearity and chaos}
\begin{figure}[H]{\label{surrohistpet}}
\centering
\includegraphics[height=.25\textheight]{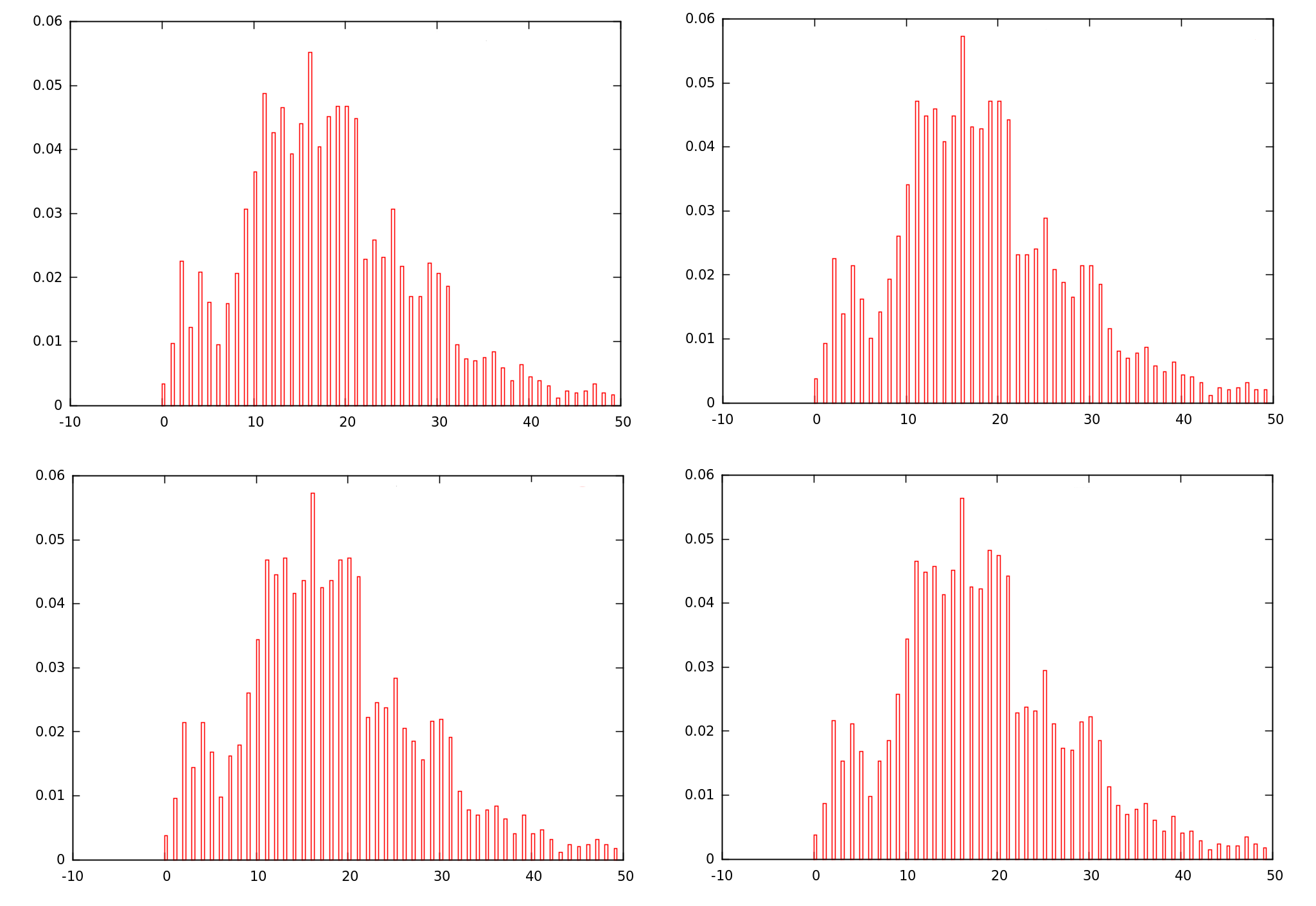}
\caption{Histogram of the number of stocks the company MINSUR, as well as his surrogate series prices}
\end{figure}

For the use of the surrogates are generated series substitute which preserve linear properties of the original series, for example, the graph(\ref{surrohistpet}) shows histograms of the original series as well as his surrogate series.

\begin{figure}[H]{\label{surropredictpet}}
\centering
\includegraphics[height=.25\textheight]{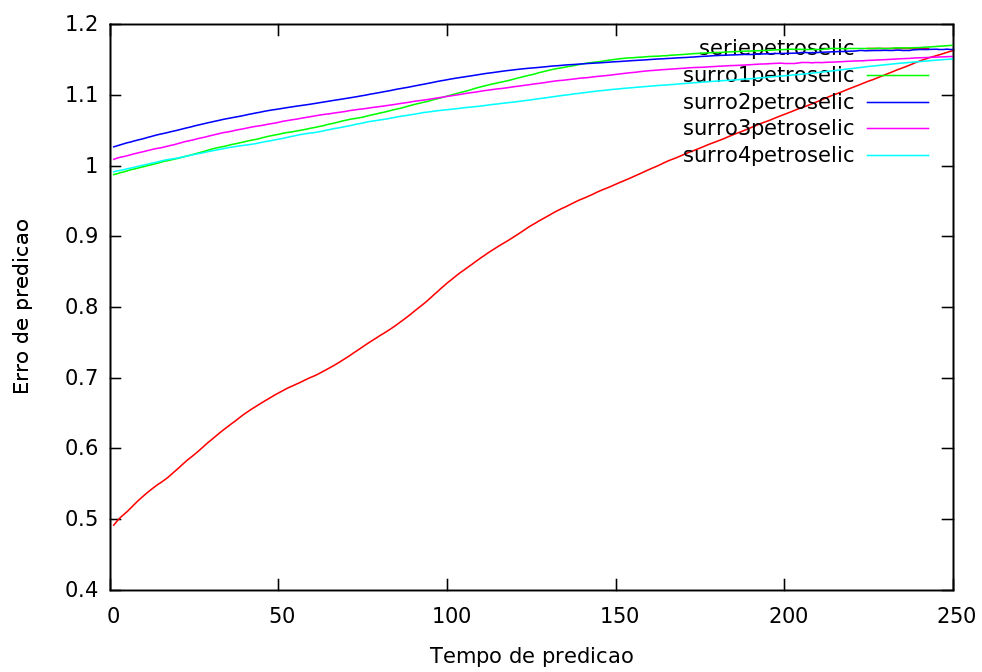}
\caption{Errors in the prediction of the original series and its surrogate series.}
\end{figure}

The next step is the calculation of the predictions of each of the series, original and substitute, and comparing related errors. The figure(\ref{surropredictpet}) shows that errors of prediction of the original series presents a relatively small short-term prediction error and substitute series shows a high prediction error from the first moment.

Therefore the null($H_0$: error of the original series prediction = surrogate series prediction error)hypothesis is rejected since the original series prediction error is less than all their substitute series prediction error. Therefore, we reject the fact that the original series is generated by a linear stochastic process.

On the other hand, the BDS test is an alternative more statistics to detect non-linear units, perform the test for several dimensions of immersion. In the chart are the values of the BDS statistics to dimensions of $1$ to $15$.
\begin{center}
\begin{tabular}{cc} \hline
Dimension of immersion(m) & BDS \\
\hline
1 & 165.270466  \\
2 & 195.653634  \\
3 & 236.554950  \\
4 & 296.226274  \\
5 & 382.825922  \\
6 & 508.150676  \\
7 & 690.060422  \\
8 & 955.415301  \\
9 & 1344.764026  \\
10 & 1919.802065  \\
11 & 2774.198843 \\
12 & 4050.599850  \\
13 & 5967.524832 \\
14 & 8860.578053 \\
15 & 13419.054424  \\ \hline
\multicolumn{2}{l}{}
\end{tabular}
\end{center}

We see the table that $|BDS|>1.96$.Therefore, we reject the null hypothesis (that the data are IID), with a degree of significance of $5\%$.

Identified the non-linearity in the series, the next step is the search for evidence of chaotic dynamics.

\subsection{The greatest exponent of Lyapunov}

 Finally, we calculate the greatest exponent of Lyapunov as an indicator of chaos. The theorem ergodic justice Oseledec method Rosenstein and Kantz to estimate the maximum exponent of Lyaunov by the formula(\ref{separaexpo}).In practice it is convenient to only calculate the maximum exponent of Lyapunov, Since the whole spectrum depends on the dimension of immersion and colors more exponents than actually exist, or less.

\begin{figure}[H]
\centering
\begin{minipage}[c]{4cm}
\includegraphics[width=5cm]{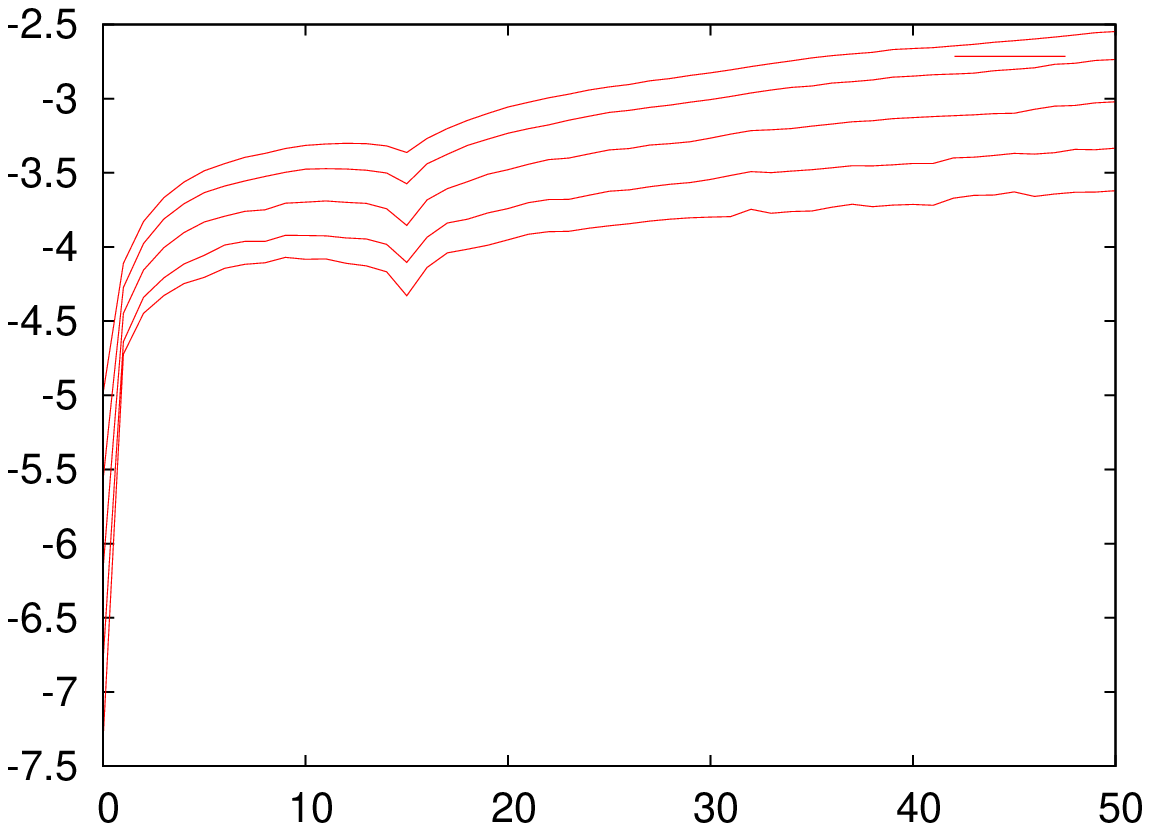}
\end{minipage}\hspace{0.1cm}
\begin{minipage}[c]{4cm}
\includegraphics[width=5cm]{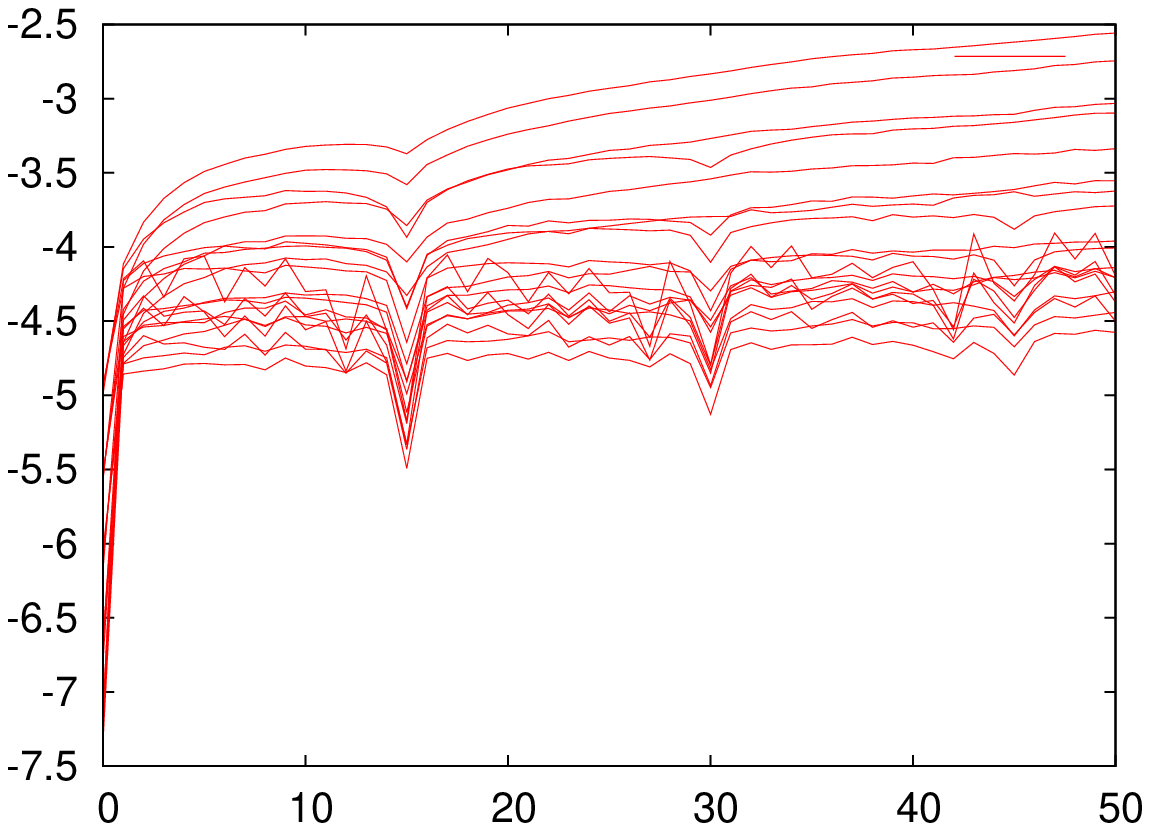}
\end{minipage}
\caption{Departure of paths using the algorithm of Rosenstein and Kantz respectively.}\label{figures}
\end{figure}

\section{Conclusions}
Using the analysis nonlinear and chaos theory applied in the series of prices of the shares of the mining company MINSUR obtained the following results: 
\begin{enumerate}
\item The data contains non-linear dependence.
\item We cannot conclude the existence of chaos in the series.
\end{enumerate}

Even considering non-linear dependence in the data results to detect chaos are not conables. Some authors\cite{Mikael} estimated the largest Lyapunov exponent using only the first iteration of estrangement. 

A job to follow is the search for presence of orbits periodically unstable, using for this purpose, the recurrence maps or maps of Poincaré, as well as the presence of a fractal attractor.

\section{Acknowledgements}
The authors are so grateful to the Department of mathematics applied of the Universidad Federal de Rio de Janeiro, UFRJ, and the Department of Sciences of the University private of the North, UPN, Trujillo-Perú, where this study was developed.


\end{document}